\documentclass{article}
\usepackage{spconf,amsmath,graphicx}

\title{Visually Guided Self Supervised Learning of Speech Representations}

\name{Abhinav Shukla$^{1}$\sthanks{Abhinav Shukla's work was supported by a PhD Scholarship by Samsung R\&D Institute, UK.},  Konstantinos Vougioukas$^{1}$,  Pingchuan Ma$^{1}$, Stavros Petridis$^{1,2}$, Maja Pantic$^{1,2}$}

\address{$^{1}$Imperial College London, UK \\
$^{2}$Samsung AI Centre, Cambridge, UK \\
\texttt{a.shukla@imperial.ac.uk, stavros.petridis04@imperial.ac.uk}}
\begin{document}
%
\maketitle
\begin{abstract}
Self supervised representation learning has recently attracted a lot of research interest for both the audio and visual modalities. However, most works typically focus on a particular modality or feature alone and there has been very limited work that studies the interaction between the two modalities for learning self supervised representations. We propose a framework for learning audio representations guided by the visual modality in the context of audiovisual speech. We employ a generative audio-to-video training scheme in which we animate a still image corresponding to a given audio clip and optimize the generated video to be as close as possible to the real video of the speech segment. Through this process, the audio encoder network learns useful speech representations that we evaluate on emotion recognition and speech recognition. We achieve state of the art results for emotion recognition and competitive results for speech recognition. This demonstrates the potential of visual supervision for learning audio representations as a novel way for self-supervised learning which has not been explored in the past. The proposed unsupervised audio features can leverage a virtually unlimited amount of training data of unlabelled audiovisual speech and have a large number of potentially promising applications.

\end{abstract}
\section{Introduction}
\label{sec:intro}

Deep neural networks trained in a supervised manner are a popular contemporary choice for various speech related tasks such as automatic speech recognition (ASR), emotion recognition and age/gender recognition. However they are a double-edged sword by virtue of providing extremely good performance given that large scale annotated data is available, which is usually expensive or time consuming. For problems like emotion recognition, reliably annotated data is also extremely scarce and even modern datasets are very limited in size. Transfer learning approaches attempt to solve this problem by domain adaptation but even they need a large amount of annotated data for the primary task and generalization is not a guarantee. Self supervision is an interesting way to attempt to combat this paucity of labeled data by capturing the intrinsic structure of the data. The idea behind self supervision is to find a `pretext task / proxy task' for the network to learn that does not require any explicit labeling, but instead the data's inherent structure \textit{provides} the labels.

There have been numerous recent works on self supervised representation learning, especially in computer vision. For example, Gidaris et. al. \cite{gidaris2018unsupervised} predict rotations for unlabeled images that have been rotated by a known amount, which drives the features to encode information about the object shape and appearance. Other works try to predict the relative location of patches \cite{doersch2015unsupervised}, temporal order of frames in a video \cite{fernando2017self}, or audio-visual synchronization \cite{korbar2018cooperative, multisensory2018}. Even in natural language processing, extremely popular recent works like ELMo \cite{peters2018deep} and BERT \cite{devlin2018bert} are based on predicting the next token of text based on the history in a self supervised way. A few works also exploit the relationship between modalities, such as by predicting cyclic transitions \cite{pham2019found}, the relationship between ambient sound and vision \cite{owens2018learning}, and cross-modal prediction based fusion \cite{petridis2015prediction}. All of these works have shown that it is possible to learn robust multi-task representations from a large amount of unlabeled data that is inexpensive to obtain.

There has also been a wave of recent work on self supervised audio-only representation learning.  CPC (Contrast Predictive Coding) \cite{oord2018representation} and APC (Autoregressive Predictive Coding) \cite{chung2019unsupervised} are similar approaches that model the next token of a speech segment given the history. Another method called LIM (Local Info Max) \cite{ravanelli2018learning} is based on maximizing the mutual information among randomly chosen windows in a recent unsupervised way to learn speaker embeddings. Wav2vec \cite{schneider2019wav2vec} is also an unsupervised pre-training method used in the context of speech recognition. Self supervised audio features have also been proposed for mobile devices \cite{tagliasacchi2019self}. Another very relevant recent work is PASE (Problem Agnostic Speech Encoder) \cite{pascual2019learning}, which aims to learn multi-task speech representations from raw audio by training an encoder to predict a number of handcrafted audio features and properties.

In this paper, we propose a self supervised way to learn multi-task speech representations by leveraging the visual modality (inspired by our prior work \cite{vougioukas2018end}). Specifically, we make the following research contributions: \textbf{\textit{(i)}} We animate a still image to generate speech video by conditioning on the corresponding audio. In doing so, the audio encoder part of our network learns useful features that are necessary to produce realistic facial and lip movements, both of which are highly correlated with the presence of emotion and that of particular phonemes. \textbf{\textit{(ii)}} These features are essentially audio only features that have been \textit{guided by} the visual modality during training, and can thus be tested even on speech datasets that do not have the visual modality.  \textbf{\textit{(iii)}} The proposed features give state of the art performance  on discrete emotion recognition on the CREMA-D \cite{cao2014crema} and Ravdess \cite{livingstone2018ryerson} datasets, and competitive performance with other self-supervised features on ASR on the GRID \cite{cooke2006audio} and SPC datasets \cite{warden2018speech}. This shows the potential of visual supervision for learning audio representations.

\section{Self Supervised Speech Representation Learning by Facial Animation}
\label{sec:method}
\begin{figure}
    \centering
    \includegraphics[width=\columnwidth]{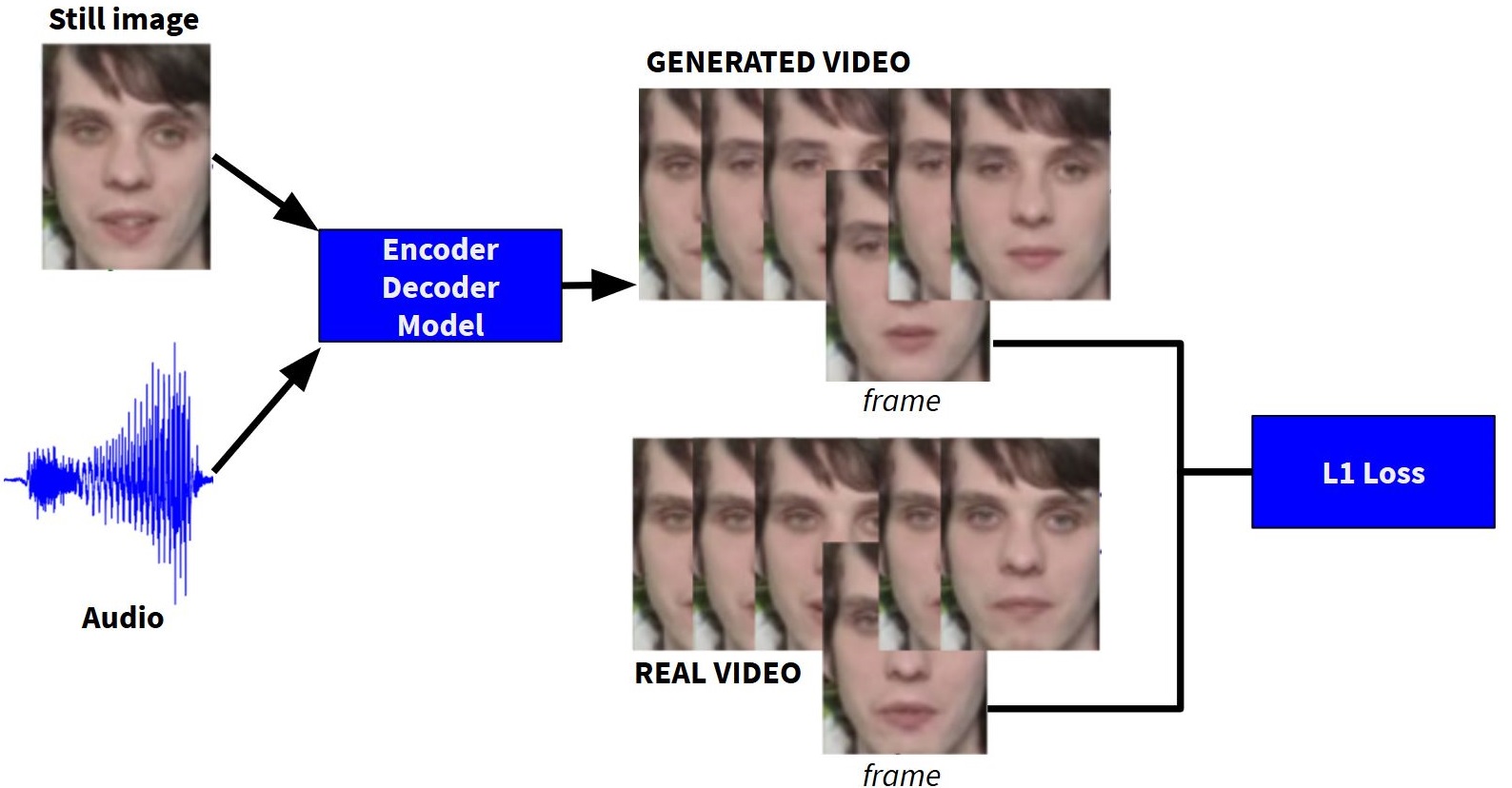}
    \caption{An overview of our proposed model. We generate a video from a still face image and the corresponding audio and optimize the reconstruction loss.}
    \label{fig:overview}
\end{figure}
\begin{figure}[t]
    \centering
    \includegraphics[width=\columnwidth]{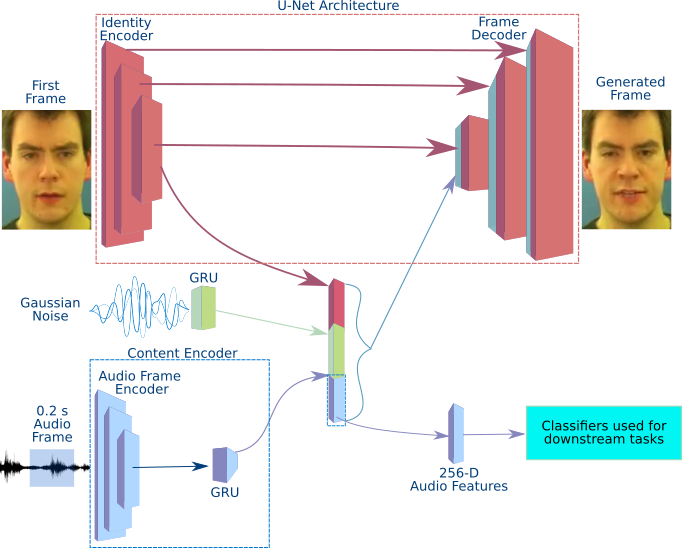} \\

    \vspace{0.2cm}

    \includegraphics[width=\columnwidth]{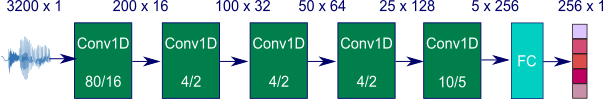}
    \caption{(Top) The architecture of the proposed model, (Bottom) the architecture of the audio encoder component which we extract features from after self supervised training.}
    \label{fig:generator}
\end{figure}

The proposed architecture is shown in Fig. \ref{fig:overview} and is based on our prior work on speech-driven facial animation \cite{vougioukas2018end}. The model is a temporal encoder-decoder which takes a still image (frame from a 25 fps video) and an audio singal as input. The audio (16 kHz waveform) is divided into overlapping windows of 200 ms, with each window centered around a video frame. The model itself can be divided into three subnetworks as shown in Fig. \ref{fig:generator}, namely the content encoder (6 layer 1D CNN audio encoder + GRU), the identity encoder (6 layer 2D CNN) and the frame decoder (U-Net \cite{ronneberger2015u} architecture with skip connections from identity encoder, layer sizes and parameters are same as U-Net).

The audio frame encoder (see bottom of Fig. \ref{fig:generator}) converts an 3200x1 audio window into a 256-dimensional feature vector $z_{aud}$ as shown. Similarly, the identity encoder, which is made of 6 (Conv2D - BatchNorm - ReLU) blocks, reduces
a 96x128 input image to a 128x1 feature vector $z_{id}$. We also use a Noise Generator capable of producing noise that is temporally coherent. A 10 dimensional vector is sampled from a Gaussian distribution with mean 0 and variance of 0.6 and passed through a single-layer GRU to produce the noise sequence. This latent representation $z_n$ accounts for randomness in the face synthesis process. The latent representation is the concatenation of $ z_{aud}, z_{id}$ and $z_n $. This embedding then goes through the frame decoder, which is a CNN that uses strided transposed convolutions to produce the video frames. The skip connections to the identity encoder help in preserving subject identity.

An L1 reconstruction loss between a random frame from the generated video and the corresponding frame from the real video is used to train the network. We use the Adam optimizer with a learning rate of 0.06 that is decayed by a factor of 0.98 every 10 epochs. The network learns to reconstruct the face. In doing so, the audio encoder is driven to produce useful speech features that contain information about mouth and facial movements. These representations can then be used for downstream tasks like ASR and emotion recognition.

\section{Datasets and Baselines}
\label{sec:majhead}

\begin{table*}[t]\centering
\small
\begin{tabular}{|c|c|c|c|c|c|c|c|c|}
\hline
\multicolumn{2}{|c|}{\textbf{Discrete Emotion Recognition}} & \multicolumn{7}{c|}{Method (Accuracy)}                                    \\ \hline
Pretrain Dataset               & Eval Dataset               & AVENet \cite{arandjelovic2018objects} & Cooperative \cite{korbar2018cooperative} & MFCC$^{\star}$ & CPC \cite{oord2018representation} & APC \cite{chung2019unsupervised} & PASE \cite{pascual2019learning} & \textbf{Ours} \\ \hline
CREMA-D                          & CREMA-D                      &  37.72      &   28.08          &  41.50    & 43.61      & 46.80    &  47.80    &   \textbf{55.01}            \\
TCD TIMIT                      & CREMA-D                      &  28.69      &   28.81          &  41.50       &  30.60   & 39.20    &  39.32    &   \textbf{49.39}            \\
LRW                            & CREMA-D                      &  29.39      &   28.53          &  41.50       &  34.31      & 41.30    &  43.16    &  \textbf{47.68}             \\ \hline
CREMA-D                          & Ravdess                    &  26.97      &   17.53          &  28.32       &  26.17     & 28.16    & 23.35     &  \textbf{41.34}            \\
TCD TIMIT                      & Ravdess                    &  20.03      &   20.72          &  28.32       &  26.01  & 32.21    & 31.76     &   \textbf{44.04}           \\
LRW                            & Ravdess                    &  19.07      &   21.38          &  28.32       &  29.05      & 34.63    & 30.05     &  \textbf{41.92}             \\ \hline
\end{tabular}

\caption{Discrete emotion recognition results (accuracy) presented on the CREMA-D (6 balanced classes, chance = 16.66) and Ravdess (8 balanced classes, chance = 12.5) datasets. All methods are pretrained on the mentioned datasets and used as feature extractors on the evaluation datasets. $^{\star}$MFCC's are used only in a supervised way on the evaluation datasets.}
\label{tab:emotion}
\end{table*}
\subsection{Datasets}
\label{ssec:datasets}
This section introduces the various audiovisual datasets that were used in the work either for pretraining or evaluating the baseline and proposed models. For all datasets, we divide the data into training, validation and test sets with all samples from each speaker belonging to a particular set only.

\begin{table}[t]
\begin{center}
\tabcolsep=0.06cm
\small
\begin{tabular}{|l|r|r|r|r|r|}
\hline
Dataset & Train & Val &  Test \\
\hline
GRID   & 31639 / 26.4 & 6999 / 5.80 & 9976 / 8.31 \\
TCD TIMIT   & 8218 / 9.10 & 686 / 0.80 &977 / 1.20 \\
LRW & 112658 / 36.3 & 5870 / 1.90& 5980 / 1.90\\
CREMA-D & 11594 / 9.70 & 819 / 0.70 & 820 / 0.68 \\
Ravdess & 1509 / 1.76 & 415 / 0.48 & 519 / 0.60 \\
SPC & 51094 / 14.2 & 6798 / 1.88 & 6835 / 1.89 \\
\hline
\end{tabular}
\vspace{-0.25cm}
\end{center}
\caption{The samples and hours (number, time) of audiovisual speech data in the training, validation and test sets of each dataset.}
\label{tab:Subjects}
\end{table}

The GRID dataset \cite{cooke2006audio} contains audio-visual speech recordings of subjects with a frontal view. It has 33 speakers, each of whom speak 1000 sentences containing six words. We use GRID as an ASR evaluation dataset. The TCD TIMIT \cite{harte2015tcd} dataset contains 59 speakers uttering 100 phonetically rich sentences sourced from the original TIMIT ASR dataset.We use TCD TIMIT as a pretraining dataset. The LRW dataset \cite{chung2016lip} is a large, in-the-wild dataset of 500 different isolated words primarily from BBC recordings. We use a subset of LRW that has only nearly frontal videos (with yaw, pitch and roll restricted to a maximum of 10 degrees). We use LRW as a large sized pretraining dataset. The CREMA-D dataset \cite{cao2014crema} contains a diverse set of 91 actors who utter 12 sentences multiples times each with a different level of intensity for each of 6 basic emotional labels (anger, fear, disgust, neutral, happy, sad). We use CREMA-D for both pretraining and evaluation for emotion recognition, but not for ASR because it is phonetically very limited even though it is larger than TCD TIMIT. The Ravdess dataset \cite{livingstone2018ryerson} contains 1440 samples of 24 different actors who acted out two sentences with 8 different basic emotions (anger, calm, sad, neutral, happy, disgusted, surprised, fear) and two different intensity levels. We use Ravdess as an emotion recognition evaluation dataset. The SPC (Speech Commands) dataset \cite{warden2018speech} contains 64,727 total utterances of 30 different words by 1,881 speakers. Table \ref{tab:Subjects} summarizes the dataset statistics.

\subsection{Baselines}
In this section, we introduce the other self supervised baseline methods that we compare our proposed method with. We chose a variety of methods that are both audio-only and audio-visual and have varying pretext tasks. When available, we use the original authors' code to evaluate the method. Note that none of these baselines require any labeled data whatsoever.

\begin{table*}[t]
\small
    \centering
    \begin{tabular}{|c|c|c|c|c|c|c|}
\hline
\multicolumn{3}{|c|}{\textbf{Speech Recognition}} & \multicolumn{4}{c|}{Method}                                    \\ \hline
Pretrain Dataset               & Eval Dataset & Metric & MFCC (Supervised)             &  CPC \cite{oord2018representation} & PASE \cite{pascual2019learning} & \textbf{Ours} \\ \hline
LRW                         & GRID               & Word Error Rate ($\downarrow$)  &  \textbf{4.7} & 10.2     & 5.8     & 11.6  \\ 
LRW                   & SPC             &   Accuracy  ($\uparrow$)  & \textbf{91.1} &   74.4       & 89.1     & 83.3  \\
\hline

\end{tabular}
    \caption{Automatic speech recognition results presented on the GRID and SPC datasets. Compared self-supervised methods are all raw audio encoders. After pretraining as mentioned,  features are input to ESPNet for ASR with a hybrid attention/CTC architecture.}
    \label{tab:speech}
\end{table*}

AVENet \cite{arandjelovic2018objects} is a two-stream audio-visual correspondence based network. One second of audio along with the middle frame of the one second segment are passed as input to the parallel streams, with a positive pair coming from the correct point in the video and a negative pair coming from a different video. The optimization is done with a contrastive loss. We use the audio stream of the network for feature extraction. Korbar et. al. \cite{korbar2018cooperative} propose an audio-visual temporal synchronization network (Cooperative) which is also a two-stream audiovisual network but has 1 second of video frames as input as opposed to a single frame in AVENet. A positive pair of audio and video samples is one that is in sync, and there are various types of out of sync negative examples in progressive order of difficulty which are optimized with a curriculum learning strategy (easy first, hard later).

Contrast Predictive Coding (CPC) \cite{oord2018representation} is a technique that tries to model a density ratio to maximize mutual information (MI) between the target signal (random raw audio window) and the context (current raw audio window). By maximizing the MI, the method extracts underlying latent variables that the two different windows have in common. Autoregressive Predictive Coding (APC) \cite{chung2019unsupervised} is similar to CPC, the key difference being that APC directly tries to predict the future part of the signal based on the history whereas CPC tries to maximize mutual information between the target (future) and the context (present). The input features for APC are 80 dimensional log mel spectrograms with a window size of 25 ms and a step size of 10 ms. The model tries to predict the log mel spectrograms for the future windows given the history. PASE \cite{pascual2019learning} is a self supervised audio encoder trained to predict various features and properties from raw audio. While predicting these multiple attributes (MFCCs, LIM, prosody etc.), the encoder learns a robust, multi-task representation for raw audio that these tasks exemplify (e.g. prosody for emotion).

\section{Experimental Setup}
\label{sec:experiments}
We evaluate all features on: (i) Discrete Emotion Recognition and (ii) Automatic Speech Recognition.

For the \textbf{emotion recognition} task, to investigate how the quality of the representations varies with different types and quantities of training data, we first perform self supervised pretraining on either: i) CREMA-D, a small but emotionally rich database, ii) TCD TIMIT, a medium sized audiovisual speech database or iii) LRW, a large audiovisual speech database. We then use these pretrained models as feature extractors on either CREMA-D or Ravdess to get features  for each method. Finally, we train a simple 2 layer LSTM on these features with the hidden size being 256. The learning rate is 0.001 and is decayed by a factor of 0.1 every 30 epochs. We train for 100 epochs and use the weights from the epoch with the best validation accuracy for evaluation. We pass the last hidden state of the LSTM to a linear layer with a size equal to the number of target classes (6 for CREMA, 8 for Ravdess) before a softmax layer with a cross entropy loss for emotion classification. This exact same process (unsupervised feature extraction + LSTM training) is performed for all the methods being compared.

For the \textbf{speech recognition} task, we choose the GRID and SPC datasets to evaluate the features. We perform self supervised pretraining for all methods on LRW, and then we use the extracted features converted to Kaldi format for ASR. We employ the ESPNet \cite{watanabe2018espnet} toolkit for the end-to-end ASR training. We use a hybrid CTC/attention based ASR model with the default ESPNet parameters with a BLSTM encoder with 320 units and location aware attention. We train the model for 15 epochs. For decoding, we use a beam search with a beam size of 20 and a CTC weight of 0.1.

\section{Results}
\label{sec:results}

For \textbf{emotion recognition} (Table \ref{tab:emotion}), irrespective of the pretraining and evaluation dataset, our method is the best performing by a significant margin. PASE is the closest competing self supervised method when evaluating on CREMA-D, and APC is the closest method on Ravdess. AVENet and Cooperative are not able to learn equally useful emotion representations, likely due to the synchronization pretext task not being the most appropriate for emotion. Our method is also the only one that outperforms the supervised MFCC baseline in every setting. Compared to other methods, it is able to learn more robust and generalizable emotion features from a variety of pretraining datasets, showing potential for using exponentially larger datasets for self supervised training. These observations indicate that our features are useful unsupervised emotion representations, likely due to our audio features being driven to capture facial expression information (which is highly correlated with acted emotion) by visual supervision.

For \textbf{speech recognition} (Table \ref{tab:speech}), we present results only on methods that are raw audio encoders for fair comparison, because methods like APC that encode MFCC's or Mel features specifically engineered for ASR are likely to be at an advantage. We use the LRW dataset for pretraining all methods evaluated for ASR. On GRID, we achieve a WER of 11.6, while PASE is the best self supervised method with a WER of 5.8. On SPC, we achieve an accuracy of 83.34 on the test set, which is again inferior to PASE with 89.1. CPC outperforms our method on GRID with a WER of 10.2, but is much worse on SPC with an accuracy of 74.37. The supervised baseline is regardless the best performing method for ASR, both on GRID (WER 4.7) and SPC (Accuracy 91.06). 

\section{Conclusion}
\label{sec:conclusion}
In this work, we present a method to learn self supervised speech representations that are guided by video generation. We evaluate the quality of the features extracted by the audio encoder to that of features extracted by other self supervised competitor methods and find that we achieve state of the art performance for discrete emotion recognition on CREMA-D and Ravdess and competitive performance for ASR on GRID and SPC. This demonstrates the potential of cross-modal supervision for learning useful representations and the proposed visually guided supervision can be easily integrated to other self-supervised approaches.
In the future, we would like to evaluate our model on naturalistic and continuous affect recognition as opposed to the acted and discrete emotion datasets in this work.
 

\section{Acknowledgements}
 We gratefully acknowledge the support of NVIDIA Corporation with the donation of the Titan V GPU used for this research and Amazon Web Services for providing computational resources for Konstantinos Vougioukas's work

\bibliographystyle{IEEEbib}
\bibliography{refs}

\begin{thebibliography}{10}

\bibitem{gidaris2018unsupervised}
S.~Gidaris, P.~Singh, and N.~Komodakis,
\newblock ``Unsupervised representation learning by predicting image
  rotations,''
\newblock {\em arXiv:1803.07728}, 2018.

\bibitem{doersch2015unsupervised}
C.~Doersch, A.~Gupta, and A.~Efros,
\newblock ``Unsupervised visual representation learning by context
  prediction,''
\newblock in {\em ICCV}, 2015, pp. 1422--1430.

\bibitem{fernando2017self}
B.~Fernando, H.~Bilen, E.~Gavves, and S.~Gould,
\newblock ``Self-supervised video representation learning with odd-one-out
  networks,''
\newblock in {\em CVPR}, 2017, pp. 3636--3645.

\bibitem{korbar2018cooperative}
B.~Korbar, D.~Tran, and L.~Torresani,
\newblock ``Cooperative learning of audio and video models from self-supervised
  synchronization,''
\newblock in {\em NeurIPS}, 2018, pp. 7763--7774.

\bibitem{multisensory2018}
A.~Owens and A.~Efros,
\newblock ``Audio-visual scene analysis with self-supervised multisensory
  features,''
\newblock {\em arXiv:1804.03641}, 2018.

\bibitem{peters2018deep}
M.~Peters, M.~Neumann, M.~Iyyer, M.and~Gardner, C.~Clark, K.~Lee, and
  L.~Zettlemoyer,
\newblock ``Deep contextualized word representations,''
\newblock {\em arXiv:1802.05365}, 2018.

\bibitem{devlin2018bert}
J.~Devlin, M.~Chang, K.~Lee, and K.~Toutanova,
\newblock ``Bert: Pre-training of deep bidirectional transformers for language
  understanding,''
\newblock {\em arXiv:1810.04805}, 2018.

\bibitem{pham2019found}
H.~Pham, P.~Liang, T.~Manzini, L.~Morency, and B.~P{\'o}czos,
\newblock ``Found in translation: Learning robust joint representations by
  cyclic translations between modalities,''
\newblock in {\em AAAI}, 2019, vol.~33, pp. 6892--6899.

\bibitem{owens2018learning}
A.~Owens, J.~Wu, J.~McDermott, W.~Freeman, and A.~Torralba,
\newblock ``Learning sight from sound: Ambient sound provides supervision for
  visual learning,''
\newblock {\em IJCV}, vol. 126, no. 10, pp. 1120--1137, 2018.

\bibitem{petridis2015prediction}
S.~Petridis and M.~Pantic,
\newblock ``Prediction-based audiovisual fusion for classification of
  non-linguistic vocalisations,''
\newblock {\em IEEE Transactions on Affective Computing}, vol. 7, no. 1, pp.
  45--58, 2015.

\bibitem{oord2018representation}
A.~Oord, Y.~Li, and O.~Vinyals,
\newblock ``Representation learning with contrastive predictive coding,''
\newblock {\em arXiv preprint arXiv:1807.03748}, 2018.

\bibitem{chung2019unsupervised}
Y.~Chung, W.~Hsu, H.~Tang, and J.~Glass,
\newblock ``An unsupervised autoregressive model for speech representation
  learning,''
\newblock {\em arXiv:1904.03240}, 2019.

\bibitem{ravanelli2018learning}
M.~Ravanelli and Y.~Bengio,
\newblock ``Learning speaker representations with mutual information,''
\newblock {\em arXiv:1812.00271}, 2018.

\bibitem{schneider2019wav2vec}
S.~Schneider, A.~Baevski, R.~Collobert, and M.~Auli,
\newblock ``wav2vec: Unsupervised pre-training for speech recognition,''
\newblock {\em arXiv:1904.05862}, 2019.

\bibitem{tagliasacchi2019self}
M.~Tagliasacchi, B.~Gfeller, F.~Quitry, and D.~Roblek,
\newblock ``Self-supervised audio representation learning for mobile devices,''
\newblock {\em arXiv:1905.11796}, 2019.

\bibitem{pascual2019learning}
S.~Pascual, M.~Ravanelli, J.~Serr{\`a}, A.~Bonafonte, and Y.~Bengio,
\newblock ``Learning problem-agnostic speech representations from multiple
  self-supervised tasks,''
\newblock {\em arXiv:1904.03416}, 2019.

\bibitem{vougioukas2018end}
K.~Vougioukas, S.~Petridis, and M.~Pantic,
\newblock ``End-to-end speech-driven facial animation with temporal gans,''
\newblock {\em arXiv:1805.09313}, 2018.

\bibitem{cao2014crema}
H.~Cao, D.~Cooper, M.~Keutmann, R.~Gur, A.~Nenkova, and R.~Verma,
\newblock ``Crema-d: Crowd-sourced emotional multimodal actors dataset,''
\newblock {\em IEEE transactions on affective computing}, vol. 5, no. 4, pp.
  377--390, 2014.

\bibitem{livingstone2018ryerson}
S.~Livingstone and F.~Russo,
\newblock ``The ryerson audio-visual database of emotional speech and song
  (ravdess): A dynamic, multimodal set of facial and vocal expressions in north
  american english,''
\newblock {\em PloS one}, vol. 13, no. 5, pp. e0196391, 2018.

\bibitem{cooke2006audio}
M.~Cooke, J.~Barker, S.~Cunningham, and X.~Shao,
\newblock ``An audio-visual corpus for speech perception and automatic speech
  recognition,''
\newblock {\em The Journal of the Acoustical Society of America}, vol. 120, no.
  5, pp. 2421--2424, 2006.

\bibitem{warden2018speech}
Pete Warden,
\newblock ``Speech commands: A dataset for limited-vocabulary speech
  recognition,''
\newblock {\em arXiv preprint arXiv:1804.03209}, 2018.

\bibitem{ronneberger2015u}
O.~Ronneberger, P.~Fischer, and T.~Brox,
\newblock ``U-net: Convolutional networks for biomedical image segmentation,''
\newblock in {\em MICCAI}, 2015, pp. 234--241.

\bibitem{arandjelovic2018objects}
R.~Arandjelovic and A.~Zisserman,
\newblock ``Objects that sound,''
\newblock in {\em ECCV}, 2018, pp. 435--451.

\bibitem{harte2015tcd}
N.~Harte and E.~Gillen,
\newblock ``Tcd-timit: An audio-visual corpus of continuous speech,''
\newblock {\em IEEE Transactions on Multimedia}, vol. 17, no. 5, pp. 603--615,
  2015.

\bibitem{chung2016lip}
J.~Chung and A.~Zisserman,
\newblock ``Lip reading in the wild,''
\newblock in {\em ACCV}, 2016.

\bibitem{watanabe2018espnet}
S.~Watanabe, T.~Hori, S.~Karita, T.~Hayashi, J.~Nishitoba, Y.~Unno, N.~Soplin,
  J.~Heymann, M.~Wiesner, N.~Chen, A.~Renduchintala, and T.~Ochiai,
\newblock ``Espnet: End-to-end speech processing toolkit,''
\newblock in {\em Interspeech}, 2018.

\end{thebibliography}
\end{document}